\documentclass{article}
\usepackage{piner}
\begin{document}

\title{Orientation and Speed of the Parsec-Scale Jet in NGC~4261 (3C~270)}

\author{B. Glenn Piner\altaffilmark{1,2}, Dayton L. Jones\altaffilmark{2}, \& 
Ann E. Wehrle\altaffilmark{2}}

\altaffiltext{1}{Department of Physics and Astronomy, Whittier College,
13406 E. Philadelphia Street, Whittier, CA 90608; gpiner@whittier.edu}

\altaffiltext{2}{Jet Propulsion Laboratory, California Institute of Technology, 4800 Oak
Grove Dr., Pasadena, CA 91109; Dayton.L.Jones@jpl.nasa.gov, Ann.E.Wehrle@jpl.nasa.gov}

\begin{abstract}
NGC~4261 (3C~270) is an elliptical galaxy containing a 300-parsec scale nuclear disk of gas and dust
imaged by the Hubble Space Telescope (HST), around a central supermassive black hole.  Previous VLBI observations of NGC~4261
revealed a gap in emission in the radio counterjet, presumably due to free-free absorption
in the inner parsec of the accretion disk.  Here we present three 8 GHz VLBA observations of
NGC~4261 that allow us to monitor the location and depth of the gap, and check for motions in
the jet and counterjet.

The separation between the 
brightest peak and the gap is stable, with an upper limit to its motion of $0.01 c$, supporting
the interpretation of the gap as absorption by an accretion disk rather than an intrinsic
jet feature.  These observations span a time of order that required for orbiting material in the disk
to transit the counterjet, so we are able to search for density changes (clumps) in the disk by monitoring
the optical depth of the gap.  The optical depth of the gap is stable to within 20\% over 5 years at $\tau=1.1\pm0.1$,
corresponding to an electron density in the disk that is constant to within 10\%.

We measure an apparent speed in the jet of 0.52$\pm$0.07$c$.  An apparent speed could not
be measured for the counterjet due to a lack of identifiable features.  From the apparent
jet speed and the jet to counterjet brightness ratio, we calculate the viewing angle of the
jet to be $63\pm3\arcdeg$ and its intrinsic speed to be $0.46\pm0.02c$.  From the inclination
and position angles of the parsec-scale radio jet and outer HST disk rotation axis we calculate a difference
between the parsec-scale radio jet and outer HST disk rotation axis of $12\pm2\arcdeg$.
Because of its well-defined HST disk and bright parsec-scale radio jet and counterjet, NGC~4261
is ideal for studying the combined disk-jet system, and this is the first case known
to us where both the inclination and position angles of both the disk and jet have been determined.

\end{abstract}


\section{Introduction}
\label{intro}
NGC~4261 is a nearby E2 galaxy associated with the low-luminosity (FR I) radio source 3C 270.
The large-scale radio structure of NGC~4261 shows two nearly symmetric kiloparsec-scale jets,
with the slightly brighter main jet lying along a position angle
\footnote{Position angle is measured from north through east.} of $-92\pm1\arcdeg$ (Birkinshaw \& Davies 1985).
The stellar rotation axis lies along position angle $153\pm4\arcdeg$ (Davies \& Birkinshaw 1986), only
$6\arcdeg$ from the projected major axis of the galaxy (PA=$159\arcdeg$).  Davies \& Birkinshaw conclude
that the intrinsic figure of the galaxy is prolate (in which case the projected minor axis
at PA=$69\arcdeg$ is not necessarily the minor axis of the 3D figure), and that the stars may
rotate about either the longest or the shortest axis of the galaxy.  
Nolthenius (1993) gives a distance of $27h^{-1}$ Mpc to NGC~4261, or 40 Mpc for an assumed
Hubble constant of 67 km s$^{-1}$ Mpc$^{-1}$.  We use this distance to NGC~4261 throughout this paper.
At a distance of 40 Mpc, 1 milliarcsecond (mas) corresponds to 0.2 pc.

The nucleus of NGC~4261 is known to contain a central black hole with a mass of $7\times10^{8}M_{\sun}$,
and a nuclear disk of gas and dust with a diameter of $\sim$300 pc, from the HST observations
of Ferrarese, Ford, \& Jaffe (1996) (see also Jaffe et al. 1993; 1996), 
with values converted to our assumed distance.
The rotation axis of the HST disk (which is interpreted as an ``outer accretion disk'' by Jaffe et al. [1993])
is inclined 64$\arcdeg$ from the line-of-sight based on isophote fitting,
and has a position angle of $-107\pm2\arcdeg$ (on the side towards the main jet) (Ferrarese et al. 1996).  
Martel et al. (1999) measure a position angle of $-112\pm5\arcdeg$ for this axis from their HST
3CR Snapshot Survey observation of NGC~4261, and de Koff et al. (2000) measure a position angle
of $-100\arcdeg$ from a ``model absorption map'' obtained from the same data.
In this paper we use the angles measured from the deeper HST exposures by Ferrarese et al. 
The lack of a relationship between the disk rotation axis and the stellar rotation axis, and the fact that the disk is
not centered on the nucleus or the isophotal center of the galaxy, led Ferrarese et al. to consider an external
origin for the HST disk material in a merger event.

The first VLBI images of this galaxy were made at frequencies of 1.6 and 8 GHz (Jones \& Wehrle 1997, hereafter JW97).
JW97 detected a parsec-scale jet and counterjet aligned with the kiloparsec-scale jets
(suggesting long-term stability of the spin axis of the central black hole), and misaligned by
14$\arcdeg$ in position angle from the HST disk rotation axis (suggesting a warping of the disk in the region close
to the central black hole).  JW97 also detected a gap in emission on the counterjet side of the core,
at a projected distance of 0.1 pc, 
which they interpreted as absorption by a small, dense, inner accretion disk.  Subsequent VLBI images at 22 and 43 GHz
(Jones et al. 2000), and higher resolution VLBI images at 1.6 and 5 GHz (Jones et al. 2001), 
confirmed free-free absorption of radio emission by a geometrically thin disk of ionized gas, and allowed 
estimates to be made of the
electron density and magnetic field of the accretion disk on sub-parsec scales.  
Absorption by neutral hydrogen has also been detected in NGC~4261 at a projected distance of $\sim$2.5 pc
from the core by van Langevelde et al. (2000).  NGC~4261 is not the only radio galaxy to show evidence for
free-free absorption on parsec scales; such evidence has also been seen in, e.g, Cen A (Jones et al. 1996),
Hydra A (Taylor 1996), Cygnus A (Krichbaum et al. 1998), NGC~1052 (Kellermann et al. 1999), and 3C 84 (Walker et al. 2000).

In this paper we present our multi-epoch VLBI observations of NGC~4261 at 8 GHz, consisting of a
single epoch from 1995 (originally presented by JW97), and two new epochs from 1999.
The motivation for performing multi-epoch observations was twofold.  First, we wished
to monitor the position and depth of the emission gap.  The position of the gap should be stable
if due to an accretion disk, and variations in the depth of the gap could allow detection of
varying densities in the accretion disk.  Second, we wanted to measure proper motions of other features
in the jet and counterjet.  Measurement of any two of jet proper motion, counterjet proper motion,
or jet to counterjet brightness ratio, allows calculation of the intrinsic jet speed and inclination
angle.  Knowledge of the radio jet inclination angle allows the complete 3-dimensional orientation
of the disk and jet to be found (as opposed to just the position angle offset discussed above).
If all three of these quantities can be measured then even more information about the source
can be obtained (including an independent measurement of the distance to the source or the Hubble constant, e.g., Taylor \&
Vermeulen [1997] and Giovannini et al. [1998]; or separate estimates for the fluid and pattern speeds in the jet, e.g.,
Taylor, Wrobel, \& Vermeulen [1998] and Cotton et al. [1999]).

\section{Observations}
\label{obs}
We observed NGC~4261 at three epochs (1995 Apr 1, 1999 Feb 26, and 1999 Oct 21) at 8.4 GHz with the
National Radio Astronomy Observatory's Very Long Baseline Array (VLBA)
\footnote{The National Radio Astronomy Observatory is a facility of the National Science
Foundation operated under cooperative agreement by Associated Universities, Inc.}.
The 1995 observation has 1.5 hours of data on-source at 8.4 GHz, while the 1999 observations each recorded
8 hours of data on-source.  During the 1999 Oct 21 observation the antenna at Brewster, WA was 
not used due to a broken azimuth bearing.
The 1995 observation recorded left circular polarization, and the two 1999 observations
recorded right circular polarization.  All of the observations recorded a bandwidth of 64 MHz.
Calibration and fringe-fitting were done with
the AIPS software package.  Images from these datasets were produced using standard CLEAN and
self-calibration procedures from the Difmap software package (Shepherd, Pearson, \& Taylor 1994).

Images obtained from the VLBI data at these epochs are shown in Figure 1.  The 1995 dataset has
been previously presented by JW97, but is shown again here for comparison with the more recent epochs.
All images have been restored with the Gaussian beam made from the uniformly weighted data from the 1999 Feb 26 epoch 
(1.7~$\times$~0.7~mas FWHM at $-2\arcdeg$) to aid in comparison.
The true restoring beams for the other epochs were 1.9~$\times$~0.8~mas at $0\arcdeg$ for 1995 Apr 1
and 2.4~$\times$~0.7~mas at $-13\arcdeg$ for 1999 Oct 21.
The VLBI images clearly show a complex parsec-scale jet (to the west, or right) and fainter counterjet
(to the east, or left), as well as the compact core which has a peak flux density of about 100 mJy beam$^{-1}$.
The peak core flux density increased by 30\% between the two 1999 observations, from
91 to 118 mJy beam$^{-1}$.  Emission from the jet and counterjet is detectable out to about
20 mas on either side of the core in these datasets.  The gap in emission just to the east of the core
that has been interpreted as absorption by an accretion disk by JW97 and Jones et al. (2000, 2001) 
is still clearly visible in the 1999 images.  The gap is even more evident in the super-resolved color image from
1999 Oct 21 that is zoomed in on the inner jet region, shown in Figure 2.  The stability of this gap is discussed in the next section.

\begin{figure*}
\plotfiddle{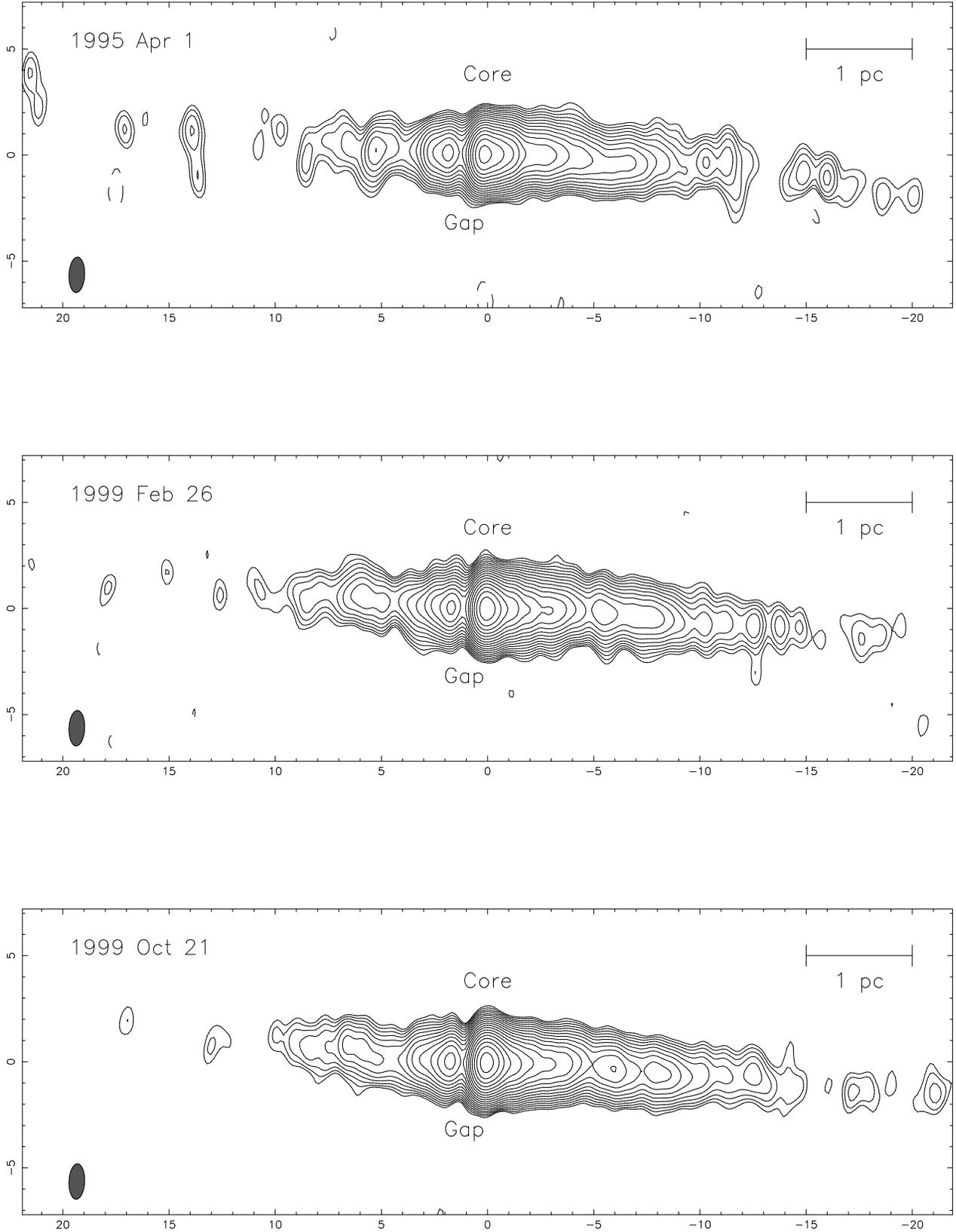}{8.5 in}{0}{90}{90}{-282}{-50}
\caption{CLEAN images of NGC~4261 from the 8.4 GHz VLBA observations on 1995 April 1, 1999
February 26, and 1999 October 21.  The axes are labeled in milliarcseconds (mas).
All images have been restored with the uniformly weighted Gaussian beam from the 1999 Feb 26 epoch 
of 1.7~$\times$~0.7~mas FWHM at $-2\arcdeg$.  The lowest contour in each image
has been set to 3 times the rms noise level in the image, and is equal to 0.32 mJy beam$^{-1}$
for the 1995 epoch and 0.19 mJy beam$^{-1}$ for the 1999 epochs.
Successive contours are each a factor of $\sqrt{2}$ higher.
The peak flux densities are 98, 91, and 118 mJy beam$^{-1}$ for the three epochs, respectively.}
\end{figure*}

\begin{figure*}
\plotfiddle{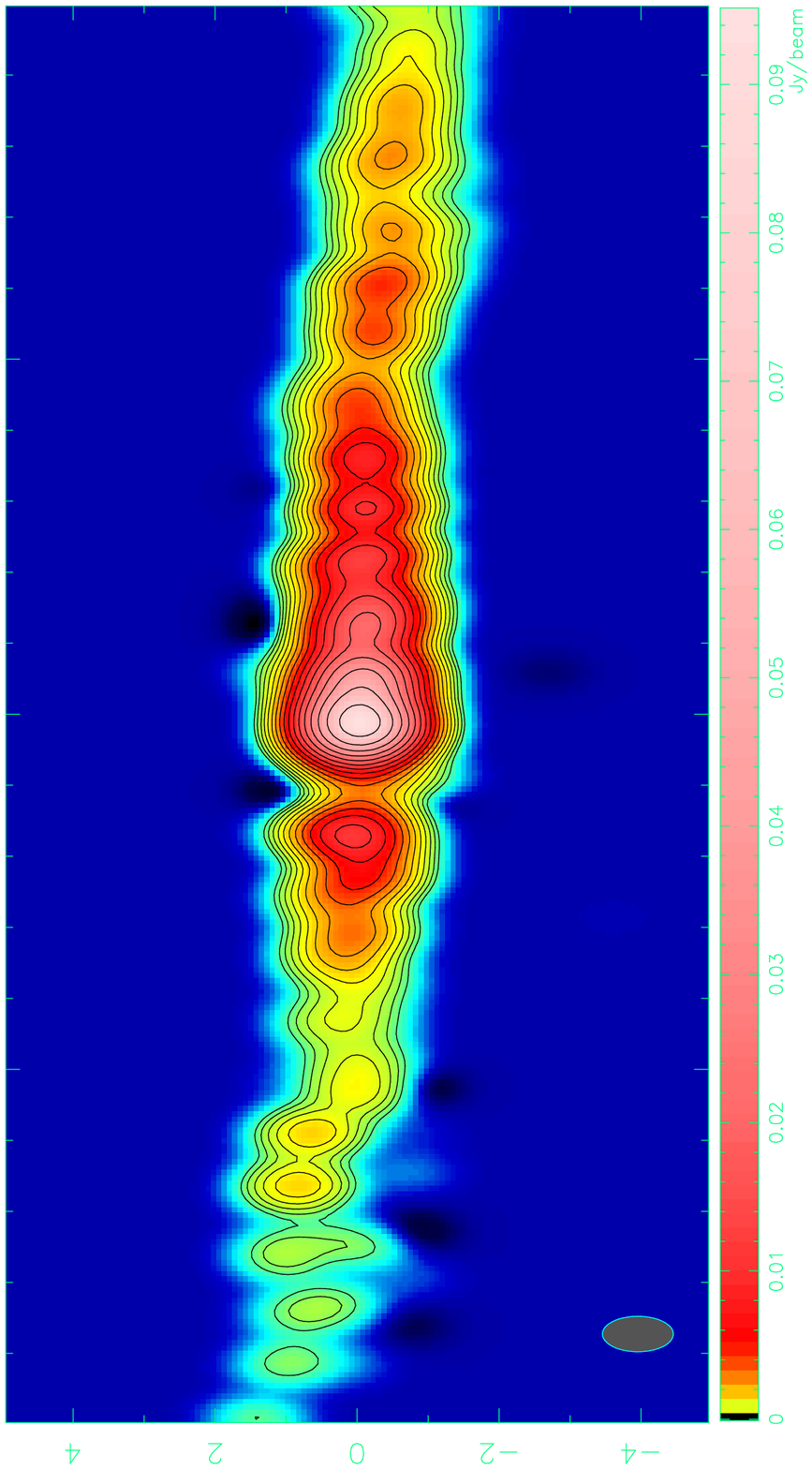}{4.0 in}{-90}{75}{75}{-310}{350}
\caption{Super-resolved false color CLEAN image of NGC~4261 from the 8.4 GHz VLBA observation
on 1999 October 21.  The displayed region extends to $\pm$10 mas from the presumed core in
right ascension, and $\pm$5 mas in declination.  The gap in emission is clearly visible 1 mas
east of the core.  The restoring beam has a FWHM of 1.0~$\times$~0.5~mas.
The lowest contour represents a flux density of 0.3 mJy beam$^{-1}$, and 
successive contours are each a factor of $\sqrt{2}$ higher.}
\end{figure*}

\section{Variability of the Jet, Counterjet, and Accretion Disk}

\subsection{Flux Density Profiles of the Jet and Counterjet}
A standard technique for analyzing VLBI images involves modeling the jet as a series of elliptical
Gaussians, by fitting these Gaussians either to the images or to the visibilities.
This works well for a jet that can be represented by a series of discrete `blobs', which
is not the case for NGC~4261.  The jets of NGC 4261, shown in Figure 1, are relatively smooth,
with modest local maxima or minima superposed on an otherwise smoothly declining jet.
Modeling the jet as a series of elliptical Gaussians would require a very large number of
components, and is not physically justified in this case.  Instead, we study
the variability of the jet and counterjet using flux density profiles along the jets.
To construct these flux density profiles we rotated the images by 3$\arcdeg$ (to take into account the
$-93\arcdeg$ position angle of the jet), restored the images with pixels one-half the size of
those used in Figure 1 (but using the same beam size) to produce smoothly varying curves, and
then summed the flux density along columns of pixels perpendicular to the jet.  This summed flux density is
expressed in units of Jy mas$^{-1}$, the total flux density is then given by integrating the area under
the flux density profile curves.

\subsection{Location and Optical Depth of the Gap}
If the gap in emission to the east of the core is due to absorption by an accretion disk,
then the location of the gap should be stable.  In this scenario, the local maximum in emission
to the east of the gap (see Figure 1) is due to the cutoff of this absorption rather than
an enhancement of the intrinsic jet emission.  This maximum should also be stable, and
should not propagate down the jet as do standard VLBI components.

Figure 3 shows flux density profiles along the first 4 mas of the counterjet at all three epochs,
where the peak of the core emission has been used to align the profile plots.
These profiles clearly show both the gap in emission and the local maximum to the east of the gap.
To check the stability of the gap location, we measured the separation between the location
of the maximum flux density in the core region and the minimum flux density in the gap region.
This separation was measured to be 1.125$\pm$0.05, 1.05$\pm$0.05, and 1.125$\pm$0.05 mas at 1995 Apr 1, 1999 Feb 26,
and 1999 Oct 21 respectively, consistent with a constant position of the gap relative to the core.
The upper limit to the change in the core-gap separation over these 4.6 years is 0.075 mas,
corresponding to a proper motion upper limit of 0.016 mas/yr, or an upper limit to the
apparent speed of this feature of 0.01$c$, which is quite slow.  The stability
of the core-gap separation is consistent with the gap 
being due to absorption by a parsec-scale accretion disk.  However, the stability is also consistent with the 
gap being a stationary feature in the jet flow.
The strongest evidence for the nature of the gap is the spectrum of the 
gap as determined from multi-frequency VLBI observations (Jones et al. 2000, 2001).

\begin{figure*}
\plotfiddle{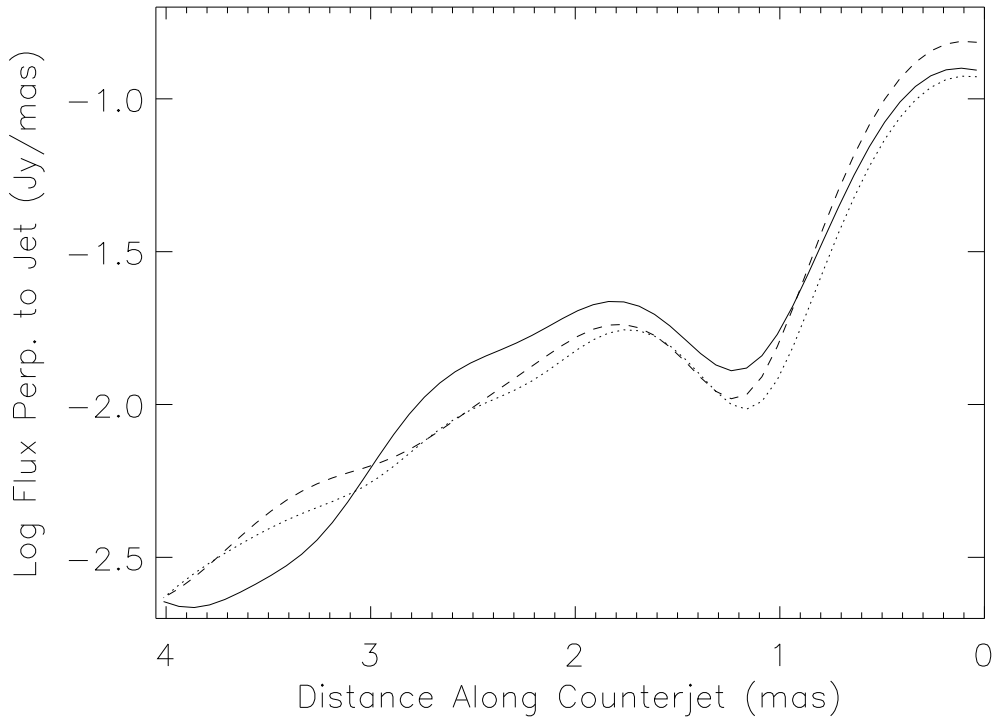}{4.0 in}{90}{60}{60}{209}{-26}
\caption{Flux density profiles along the counterjet of NGC~4261.  Epoch 1995 Apr 1 is represented
by the solid line, 1999 Feb 26 by the dotted line, and 1999 Oct 21 by the dashed line.}
\end{figure*}

If the material in the accretion disk is clumpy then we might expect to see variations in the
depth of the gap as regions of the accretion disk of varying density pass in front of the counterjet.
What would be the expected timescale for such variations?  Given an inclination angle of $64\arcdeg$
for the disk rotation axis (Ferrarese et al. 1996), the gap at a projected distance of 1.1 mas along the
counterjet corresponds to a radius of 2.5 mas, or 0.5 pc, in the accretion disk.
The orbital period at a radius of 0.5 pc about a black hole of $7\times10^{8}M_{\sun}$
(Ferrarese et al. 1996) is $\approx10^{3}$ years.  From Jones et al. (2000), the full apparent opening
angle of the jet is between 0.3$\arcdeg$ and 5$\arcdeg$.  Using these two limits, the time for
a given piece of material in the disk to pass in front of the counterjet will be of order 1-10 years.
Thus, if there are density variations, we may be able to see them over the timescale of the observations
reported here.  Our observations would be sensitive to clumps with a size of order the width of the
counterjet at the gap location ($\sim10^{-2}$ to $10^{-3}$ parsecs from the angular limits given above) 
in the direction transverse to the counterjet, but in the direction along the counterjet would
be limited by the size of the beam to clumps with a size of order 0.1 pc, which would take
considerably longer (30 years) to completely pass over the counterjet.
A clump with these dimensions, and with the path lengths and number densities given by Jones et al. (2001),
would have a mass of order $10^{-2}$ to $10^{-1} M_{\sun}$.

The optical depth in the region of the gap can be estimated by interpolating the jet emission
over the gap region, and then comparing this estimated intrinsic flux density
with the observed flux density.
We assume a power-law form for the intrinsic fading of the counterjet with distance from
the core (see JW97 and Xu et al. 2000), and find the power-law index by fitting the counterjet
flux density profiles to the east of the gap, between 2 and 9 mas from the core.
We find power-law indices of $-$2.0 for the 1995 epoch, and $-$1.8 for the two 1999 epochs.
The optical depths estimated in this fashion are 1.1$\pm$0.1, 1.2$\pm$0.1, and 1.0$\pm$0.1
for 1995 Apr 1, 1999 Feb 26, and 1999 Oct 21 respectively.  Since the optical depth is determined from
flux density ratios, it should be unaffected by changes in the absolute amplitude scales, and we expect the main source of error
in the optical depth measurement will come from imaging errors.  Errors on the optical depth were estimated
by having two of the authors independently image each dataset, and comparing the optical depths
calculated from these independently imaged datasets.  The optical depth measurements indicate that no
variations in optical depth have been detected to within about 20\%.  Since for free-free absorption
the optical depth is proportional to the square of the electron density, this corresponds to
an electron density that is constant to within about 10\% (assuming the path length through  
the disk has remained constant).  Evidently, the accretion disk
material in NGC~4261 is relatively smooth, at least on the scales sampled by these observations.

\subsection{Jet Proper Motion}
Since elliptical Gaussian components do not provide a good model for the NGC~4261 jet,
we instead made estimates of jet proper motion from motion of local maxima 
on the jet profile plots and on the VLBI images.  Figure 4 shows flux density profiles along
the jet for the two 1999 epochs.  Since even a relatively slow apparent speed would
cause a motion of several milliarcseconds between 1995 and 1999, connecting features between the 1995
epoch and the 1999 epochs is not feasible, and we look for proper motion only between the
two 1999 epochs.  For the jet, two well-defined local maxima are located at about 6
and 8 mas from the core in 1999, and they show a clear movement of about 0.5 mas
between the two 1999 epochs, using the peak of the core to align the profiles. 
The movement of the first of these local maxima is marked on Figure 4.
These local maxima are also clearly visible on the VLBI images from 1999 (in particular 1999 Oct 21,
bottom panel of Figure 1, 6 and 8 mas west of the core respectively).
By measuring the shift of these features between the VLBI images, as well as on the jet profile plots
(using both direct measurement and an algorithm that shifts one curve with respect to the other
until the sum of the squares of the difference between the two curves is minimized),
we measure an apparent jet proper motion of 0.83$\pm$0.11 mas yr$^{-1}$.
This corresponds to an apparent speed of 0.52$\pm$0.07$c$.

\begin{figure*}
\plotfiddle{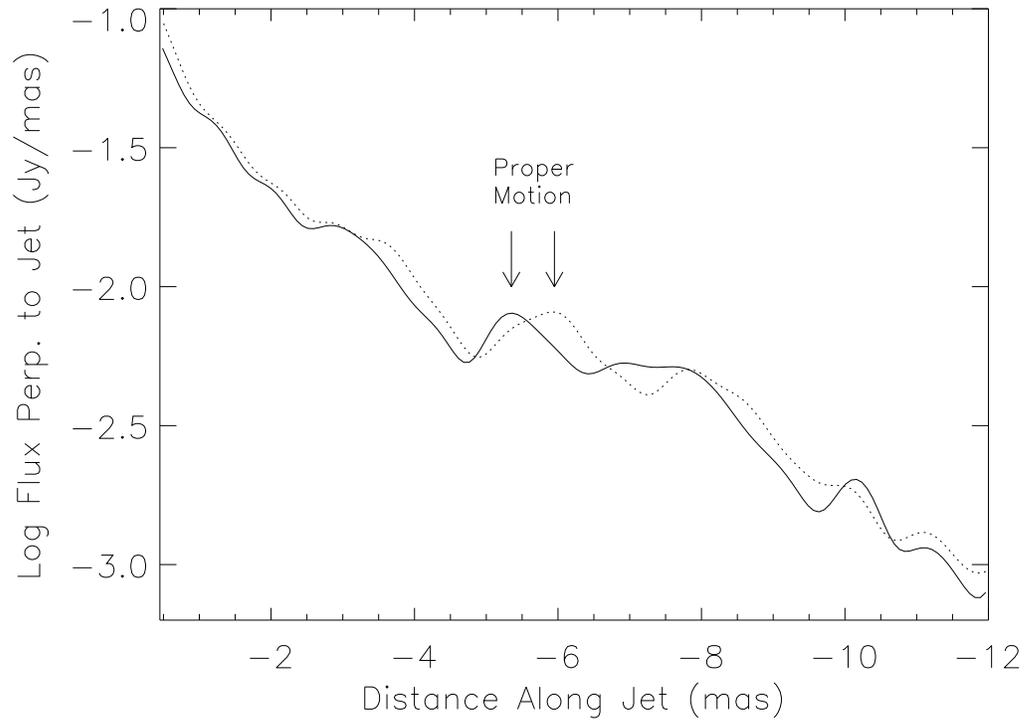}{4.0 in}{90}{60}{60}{209}{-26}
\caption{Flux density profiles along the jet of NGC~4261.  Epoch 1999 Feb 26 is represented
by the solid line, 1999 Oct 21 by the dotted line.}
\end{figure*}

For the counterjet, any such local maxima
that are visible during 1999 are at flux density levels much lower
than those in the jet, and we conclude that an apparent
speed for the counterjet cannot be reliably measured from
these two epochs.  Measurement of proper motion in the counterjet might be possible
with additional epochs at a somewhat higher sensitivity than these, and would
be valuable in constraining the properties of this source.

\section{Orientation of the Jet, Counterjet, and Accretion Disk}
\subsection{Radio Jet Position Angle}
The jets of NGC~4261 are predominantly straight, as can be seen from the plot of the jet ridgeline
from the 1999 Oct 21 observation shown in Figure 5.  Small bends of a few degrees are present, however.
The jet begins at a position angle close to $-90\arcdeg$, and at five mas from the
core bends slightly to the south.  Similar (but inverted) behavior is seen on the counterjet side.
In what follows we use the mean position angle of the VLBI jet over the inner 12 mas of
$-93\pm1\arcdeg$, measured from plots of the jet ridgeline, as the VLBI position angle.
This agrees within the errors with the kiloparsec-scale jet position angle
of $-92\pm1\arcdeg$ (Birkinshaw \& Davies 1985).

\begin{figure*}
\plotfiddle{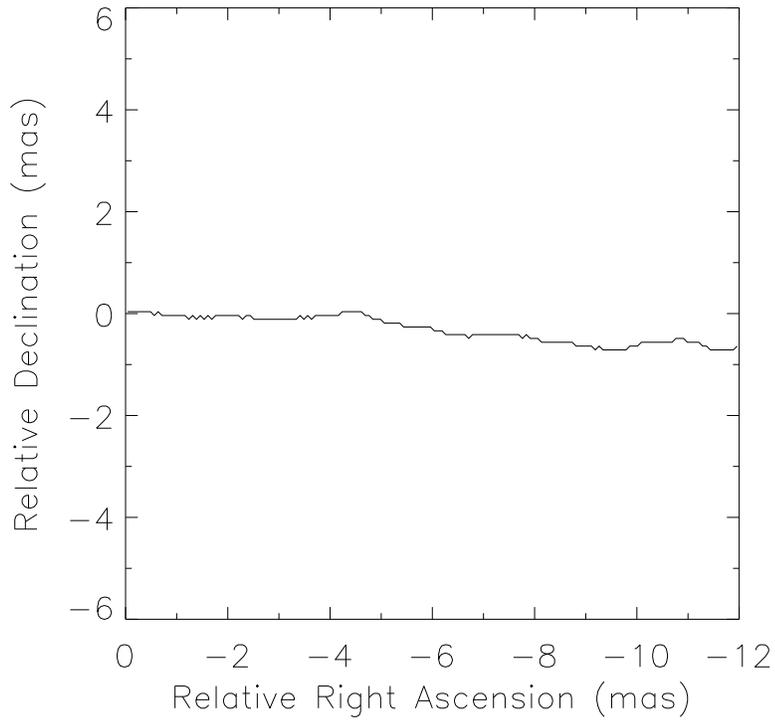}{4.0 in}{0}{60}{60}{-200}{-76}
\caption{Ridgeline of the jet from the 1999 Oct 21 observation.
The mean position angle is $-93\pm1\arcdeg$.}
\end{figure*}

\subsection{Radio Jet Inclination and Speed}
\label{inclination}
The intrinsic speed of the jet $\beta$ and the
inclination angle of the jet $\theta$ can be obtained from any two of the following
three observable quantities: the apparent speed of the jet $\beta_{app}$, the apparent
speed of the counterjet, or the jet to counterjet brightness ratio $J$.  In this section
we use measured values for $\beta_{app}$ and $J$ to calculate the intrinsic speed and inclination
angle of the jet.  The apparent jet speed and the jet to counterjet brightness ratio are given
in terms of the intrinsic speed and inclination angle of the jet by (see e.g., Ghisellini et al. 1993)
\begin{equation}
\beta_{app}=\frac{\beta\sin\theta}{1-\beta\cos\theta}
\label{betaapp}
\end{equation}
and
\begin{equation}
J=\left(\frac{1+\beta\cos\theta}{1-\beta\cos\theta}\right)^{p},
\label{ratio}
\end{equation}
where $p=2-\alpha$ for a continuous jet, and $\alpha$ is the spectral index expressed as $S\propto\nu^{\alpha}$.
Solving for $\beta$ and $\theta$ from equations (\ref{betaapp}) and (\ref{ratio}) yields
expressions for these quantities in terms of the observable quantities $\beta_{app}$ and $J$:
\begin{equation}
\theta=\arctan\frac{2\beta_{app}}{J^{1/p}-1}
\label{theta}
\end{equation}
and
\begin{equation}
\beta=\frac{\beta_{app}}{\beta_{app}\cos\theta+\sin\theta}
\label{beta1}
\end{equation}
or
\begin{equation}
\beta=\frac{J^{1/p}-1}{(J^{1/p}+1)\cos\theta}.
\label{beta2}
\end{equation}
These calculations assume that the jet and counterjet are intrinsically identical, 
that the bulk fluid speed in the
jet is the same as the pattern speed measured on the VLBI images, and that the distance to the source is known.
Measurement of the apparent speed of the counterjet would be valuable because it would allow any one of these three
assumptions to be tested (while still assuming the other two).  Equation~(\ref{ratio}) also assumes that
the jet's magnetic field has a random orientation, see Giovannini et al. (1994) for a discussion of this assumption.

The jet to counterjet brightness ratio was calculated by measuring the total flux density in the jet and the counterjet
for regions beyond 2 mas from the core (to avoid the gap region where the counterjet emission is absorbed)
to the visible edge of the counterjet at 9 mas.  
This assumes the jets are steady state, otherwise differences in
light travel time would need to be taken into account.  The flux density profiles support this
assumption, any `components' appear to be small perturbations on an otherwise smoothly declining jet.
The jet to counterjet brightness ratio for NGC~4261 at 8 GHz is $J=2.66\pm0.02$,
where the error was estimated from the (surely fortuitously small)
scatter in the values of $J$ measured at all three epochs.
This is a small sidedness ratio compared to typical lower limits found in a sample
of 17 FR I's by Xu et al. (2000). 

The spectral index of the jet emission is also needed.  To calculate $\alpha$, we measured the total
flux density of the jet at four frequencies (not simultaneous): 5, 8, 22, and 43 GHz.  The 5 GHz data and the 22 and 43 GHz data have
been previously presented by Jones et al. (2001) and Jones et al. (2000).  Figure 6 shows the
spectrum of the jet from data at these four frequencies, with a power law fit to the emission between 5 and
22 GHz.  At 43 GHz the jet emission is resolved out, and this point is not included in the fit.  The emission
between 5 and 22 GHz is well fit by a power law with slope $\alpha=-0.29\pm0.07$.  Error bars on the flux densities
were estimated to be 5\% at 5 and 8 GHz, and 10\% at 22 and 43 GHz, from typical antenna gain corrections given
by amplitude self-calibration.  From these measured values of apparent jet speed,
jet to counterjet brightness ratio, and spectral index, we calculate the intrinsic speed
and inclination angle of the jet from equations (\ref{theta}) to (\ref{beta2})
to be $\theta=63\pm3\arcdeg$ and $\beta=0.46\pm0.02c$.  The systematic errors resulting from
the assumptions underlying equations (\ref{theta}) to (\ref{beta2}) almost surely make the true
error in these derived quantities larger than the formal errors given above.

\begin{figure*}
\plotfiddle{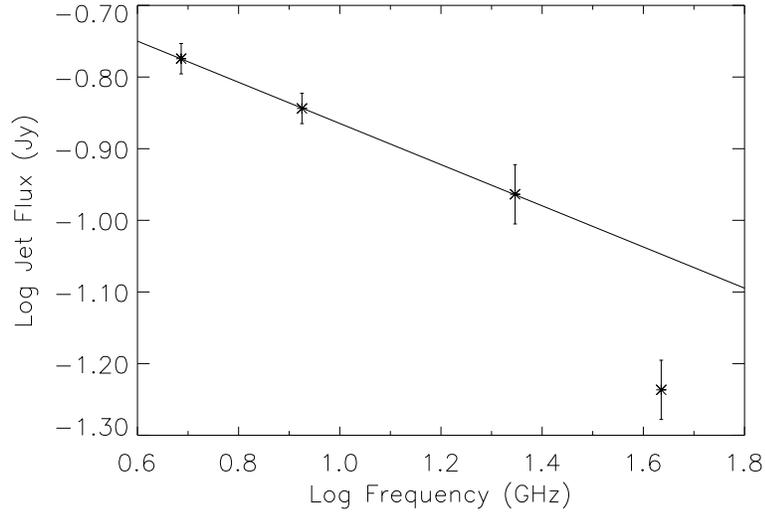}{3.0 in}{0}{60}{60}{-200}{-206}
\caption{Spectrum of the NGC~4261 parsec-scale jet.  The jet emission is resolved out at 43 GHz,
and this point is not included in the power-law fit.}
\end{figure*}

The jet speed of 0.5$c$ is only mildly relativistic, and should be compared with other known speeds in
FR I radio galaxies.  Summaries of measured FR I speeds are given by, e.g., Tingay et al. (1998)
and Cotton et al. (1999), and it is clear that a speed of 0.5$c$ is at least typical of a pattern
speed in an FR I jet, although there is evidence that bulk speeds may be larger than 
observed pattern speeds in some sources (Tingay et al. 1998).  Cotton et al. (1999) also find evidence
that some jets may accelerate from mildly relativistic speeds as they move farther out on parsec scales.
Xu et al. (2000) found that a Lorentz factor distribution with a mean of 5 reproduced the number of
twin-jet sources they observed in VLBI observations of a sample of FR I's, although the standard
deviation of the Lorentz factor distribution was not well constrained by their observations.
If the bulk flow speed in NGC~4261 were highly relativistic this would place the radio jets nearly
in the plane of the sky (in order to reproduce the low jet to counterjet brightness ratio), and would
make the misalignment between the VLBI jet and HST disk rotation axis more severe.

\subsection{Relative Orientation of the Parsec-Scale Jet and Outer Accretion Disk}
\label{relorient}
We can now compare the orientations of the VLBI jet and the HST outer disk rotation axis.
The VLBI jet has an inclination angle of $63\pm3\arcdeg$, and a position angle of $-93\pm1\arcdeg$.
The HST outer disk rotation axis has an inclination angle
of $64\pm5\arcdeg$ (from isophote fitting) and a position angle
of $-107\pm2\arcdeg$ (Ferrarese et al. 1996).  The angular offset between the
radio jet and outer HST disk rotation axis is evidently only in position angle, as the
inclination angles are identical within the errors.
The true angular separation between the VLBI jet and HST outer disk rotation axis is
then $12\pm2\arcdeg$ (the apparent separation in position angle must be multiplied by the sine of the inclination angle).
A sketch showing the position angles of the disk-jet system is shown in Figure 7.
This angular separation is consistent with the results found by de Koff et al. (2000) from the
HST 3CR Snapshot Survey that radio galaxies have jets nearly parallel to
their dust disk axes, but with some dispersion in the relation.  de Koff et al. concluded
their observed distribution of position angle differences could be reproduced by an
intrinsically flat distribution of position angle differences of the radio jet with respect
to the dust disk rotation axis between $0\arcdeg$ and $35\arcdeg$.

\begin{figure*}
\plotfiddle{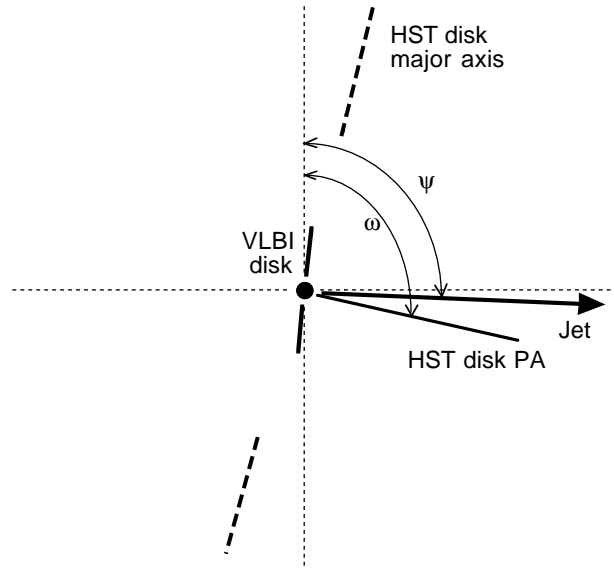}{3.5 in}{0}{60}{60}{-182}{-150}
\caption{Cartoon showing the position angles of the HST disk and VLBI jet.
Angles on this figure are not exact, and the figure is not to scale.
The VLBI-scale accretion disk is assumed to be perpendicular to the VLBI jet on this figure.
$\Psi$ represents the position angle of the VLBI jet ($-93\arcdeg$), and $\omega$ represents the
position angle of the HST disk rotation axis ($-107\arcdeg$).}  
\end{figure*}

If the VLBI jet is directed along the rotation axis of the central black hole (e.g., Rees 1978),
then the rotation axes of the central black hole and outer HST accretion disk are misaligned
by $12\pm2\arcdeg$.  The innermost regions of the accretion disk must have a rotation axis aligned with
that of the central black hole by the Bardeen-Petterson effect (Bardeen \& Petterson 1975),
but this happens close to the central black hole, at about 100 Schwarzschild radii ($R_{s}$)
(Natarajan \& Pringle 1998).  For comparison, the disk observed in absorption in this paper
is located at about $10^{4}R_{s}$, and the outer HST disk has a size of about $10^{7}R_{s}$.
Natarajan \& Pringle (1998) find that the torque that aligns the inner disk with the hole should also
align the spin of the hole with the outer accretion disk on a relatively short timescale.
However, more general calculations by Natarajan \& Armitage (1999) show that a misalignment
like that observed here can be caused by two effects: (1) the outer disk
may retain a modest warp for a long period following alignment of the hole, and (2) for holes accreting at low rates
relative to the Eddington limit (which should be the case for NGC~4261, based on its low luminosity),
the time-scale for alignment can be much longer.  

A warped accretion disk about $5\times10^{4}R_{s}$ or 0.2 pc in radius has been clearly
observed in the peculiar spiral galaxy NGC~4258; this warp has been successfully modeled with a radiation-driven warping
mechanism by Maloney, Begelman, \& Pringle (1996).  A possible warp in the disk of NGC~4261 is seen 
directly in the HST observations by Ferrarese et al. (1996) --- with FOS data they detect an `inner' disk (about half
the size of their `outer' disk) that has a rotation axis at a position angle of $-71\pm7\arcdeg$ and an
inclination of $69\pm6\arcdeg$.  Although the error on the position angle of this inner disk is
relatively large, the difference in position angle from the VLBI jet is significant, 
implying the warped disk has at least two bends.  The alternative to a warped disk is that
the current low accretion rate in NGC~4261 has been insufficient to perfectly align the angular momentum
of the central hole.  The straightness of the parsec to kiloparsec scale radio jets implies
that the orientation of the black hole angular momentum vector
has been constant for at least $10^{6}$ years (Jones et al. 2000).

\section{Conclusions}
Our new multi-epoch 8 GHz VLBA observations of NGC~4261 support the interpretation that the
gap in emission in the radio counterjet is due to absorption by a parsec-scale accretion disk.
The gap is stationary with respect to the core, with an upper limit to any apparent motion of $0.01c$.  The time required
for orbiting material in the parsec-scale disk to transit the counterjet (as seen from our vantage point) is of
order 1 to 10 years, so we have been able to use the 5-year span of VLBA observations to check 
for density changes in the accretion disk.  The optical depth ($\tau=1.1\pm0.1$) measured in the gap has been
constant over 5 years to within our errors (20\%), corresponding to an upper limit of 10\% to any changes
in electron density.  We have measured the apparent speed of the radio jet (0.52$\pm$0.07$c$), and have used
this and the jet to counterjet brightness ratio to calculate the viewing angle ($63\pm3\arcdeg$) and intrinsic
speed ($0.46\pm0.02c$) of the radio jet.  The radio jet is offset from the outer HST disk rotation axis by
$12\pm2\arcdeg$, suggesting a modest warp in the outer disk, or an inability of the current accretion
rate to perfectly align the angular momentum of the central hole.
The technique pioneered here of searching for sub-parsec scale density variations in the accretion
disk by monitoring the absorption of counterjet emission, while still limited by the resolution
and sensitivity of the VLBI observations, holds great promise for mapping accretion disk structure
on scales much smaller than those that can be directly imaged.

\acknowledgements
Part of the work described in this paper has been carried out at the Jet
Propulsion Laboratory, California Institute of Technology, under
contract with the National Aeronautics and Space Administration.
The National Radio Astronomy Observatory is a facility of the National Science
Foundation operated under cooperative agreement by Associated Universities, Inc.
BGP acknowledges support
from Whittier College's Newsom Endowment.

\end{document}